\documentclass[apj]{emulateapj}
\usepackage{apjfonts}
\usepackage{graphicx}

\submitted{Submitted to The Astrophysical Journal, August 27, 2005}
\shorttitle{ULX Sources from Accreting Binaries With IMBH?}
\shortauthors{Blecha et al.}

\begin{document}

  \title{Close Binary Interactions of Intermediate-Mass Black Holes: Possible Ultra-Luminous X-Ray Sources?}

\author{L.\ Blecha\altaffilmark{1}, N.\ Ivanova\altaffilmark{1}, V.\ Kalogera\altaffilmark{1}, K.\ Belczynski\altaffilmark{2}, J.\ Fregeau\altaffilmark{1}, and F.\ Rasio\altaffilmark{1}}
\altaffiltext{1}{Northwestern University, Dept of Physics \& Astronomy, Evanston, IL 60208\\ l-blecha@alumni.northwestern.edu; nata,vicky,fregeau,rasio@northwestern.edu}
\altaffiltext{2}{Tombaugh Fellow, Department of Astronomy, New Mexico State University, Las Cruces, NM 88003 \\ kbelczyn@nmsu.edu}

\begin{abstract}
While many observed ultra-luminous X-ray sources (ULXs, L$_{\rm X}\geq 10^{39}$ erg s$^{-1}$) could be extragalactic X-ray binaries (XRBs) emitting close to the Eddington limit, the highest-luminosity ULXs  (L$_{\rm X}>3\times 10^{39}$ erg s$^{-1}$) exceed the isotropic Eddington luminosity for even high-stellar-mass accreting black hole XRBs.  It has been suggested that these highest-luminosity ULXs may contain accreting intermediate-mass black hole (IMBH) binaries.  We consider this hypothesis for dense, young ($\sim 10^8$\,yr) stellar clusters where we assume that a 50-500 M$_{\odot}$ central IMBH has formed through runaway growth of a massive star.  Using numerical simulations of the dynamics and evolution of the central black hole's captured companions, we obtain estimates of the incidence of mass transfer phases and possible ULX activity throughout the IMBH's evolutionary history.  We find that, although it is common for the central black hole to acquire binary companions, there is a very low probability that these interacting binaries will become observable ULX sources.
\end{abstract}

 %\clearpage

\section{Introduction}

%\subsection{ULX Observations}
Extragalactic observations with the {\it Einstein} space X-ray telescope revealed a new category of sources, currently known as ultra-luminous X-ray sources (ULXs, Fabbiano 1989).  In 1990, surveys conducted with the {\it ROSAT} satellite were able to resolve many of these sources as distinct objects not associated with emission from galactic nuclei.  Up to half of the galaxies surveyed were found to contain off-nuclear ULXs \citep{cm99, rw00, llj00, cp02, cm04}.  In analogy with X-ray sources observed in our own galaxy, many ULXs could be extragalactic X-ray binaries (XRBs) with stellar-mass black hole (BH) accretors (e.g., Rappaport, Podsiadlowski, \& Pfahl 2005).  However, in many cases their luminosities exceed the Eddington limit for isotropic emission from an accreting stellar-mass BH.  The lowest luminosity that commonly defines a ULX, L$_{\rm X} \geq 10^{39}$ erg s$^{-1}$, corresponds to an Eddington black hole mass $\ga 8 {\rm M}_{\odot}$ -- still within the BH mass range allowed by stellar evolution models.  Based on the X-ray survey conducted by \citet{cp02}, ULXs with L$_{\rm X} \geq 10^{39}$ erg s$^{-1}$ are estimated to exist in $\sim12\%$ of all galaxies \citep{pc04}.  Of the 87 X-ray (2-10 keV) sources observed, 45 had L$_{\rm X} > 3\times10^{39}$ erg s$^{-1}$.  This corresponds to the Eddington luminosity for the highest-mass black holes that can form through single-star collapse, $\sim25 {\rm M}_{\odot}$ for a metallicity of $Z = 0.001$ \citep{bel04}.  We see that as much as half of this observed ULX population requires an alternative to the model of isotropic emission from a stellar-mass XRB.  

%\subsection{Evidence for IMBHs}
Observations increasingly support the idea that some ULX sources may harbor intermediate-mass black holes (IMBHs, Colbert \& Mushotzky 1999).  Until recently, no evidence existed for BHs in the wide mass range between stellar-mass and supermassive BHs.  Many spectral and timing observations conducted with {\it Chandra, ASCA,} and {\it XMM-Newton} have confirmed that ULXs are not associated with supernovae and are consistent with accreting binaries \citep{fw03}.  Additionally, ULX positions are strongly correlated with young, star-forming regions.  Examples include the nine ULXs found in the Antennae \citep{fzm01} and the highest-luminosity ULX yet observed, in M82 (peak L$_{\rm X}\sim9\times10^{40}$ erg s$^{-1}$, corresponding to an IMBH mass $\simeq 700 {\rm M}_{\odot}$ if L$_{\rm X} \simeq$ L$_{\rm X}^{\rm Edd}$; Kaaret et al. 2001; Matsumoto et al. 2001).  The young, dense clusters where ULXs are often observed are also likely sites for IMBH formation \citep{fgr05,frb05,gfr04}. 

Accretion disk spectra provide a further test for the presence of IMBHs: higher-mass BHs are expected to have cooler accretion disks modeled as multi-color disks \citep{mit84}.  Some evidence for low-temperature disks was recently found using high-resolution {\it XMM} spectra of numerous ULXs, including those in NGC 1313, NGC 4559, Ho IX, and the Antennae \citep{mill03,mill04a,mill04b}.  As \citet{cm04} note, the sources observed to have cool disks are also the ones with luminosities too high to be explained by XRBs, even if they are beamed XRBs.  Nevertheless, we note that X-ray spectral interpretation is subject to a number of model assumptions and may not be unique.

%\subsection{Beamed XRBs?}
Anisotropic, or beamed, disk emission is another possible explanation for apparently super-Eddington X-ray luminosities.  The degree to which emission is beamed can be defined by the beaming fraction b $= \Omega / 4\pi$, where $\Omega$ is the solid angle of emission.  ULXs containing beamed stellar-mass XRBs could be a short phase of rapid mass transfer in the lifetime of ordinary XRBs, such as thermal-timescale mass transfer \citep{king01}.  As demonstrated by \citet{king01}, by introducing mild beaming, the XRB falls within the stellar-mass BH range, assuming L$_{\rm X} < 10^{40}$ erg s$^{-1}$.  However, about 5-10$\%$ of the total ULX population is observed to have L$_{\rm X} \geq 10^{40}$ erg s$^{-1}$ \citep{cm04,pc04,cp02}.  At these luminosities, very massive BHs ($\sim25 {\rm M}_{\odot}$) could fall below the Eddington limit with b $\la 0.25$.  To have the required mass ratio $q = {\rm M_2/M_1} > 1$, where M$_1$ is the BH mass, these BHs would require a very massive companion.  More moderate-mass BHs would require more severe beaming fractions that are not easily justified.  Furthermore, in at least two instances observations seem to indicate isotropic X-ray emission.  These include an isotropic nebula observed arond the ULX in Ho II \citep{pk01} and the discovery of quasi-periodic oscillations (QPOs) from the brightest ULX in M82 \citep{sm03}.

%\subsection{The Role of Transiency}
Transient behavior of a ULX source is an important test for the presence of an IMBH \citep{kal04}.  If the ULX is a beamed stellar-mass XRB in thermal-timescale mass transfer, it is expected to produce persistent emission at or above the Eddington luminosity \citep{king01}.  A persistent IMBH ULX similarly would require a sustained $\dot {\rm M}$ comparable to the Eddington mass transfer rate:
\begin{equation}
 \dot {\rm M}_{\rm Edd} \simeq 2.3\times10^{-9} \left( {{\rm M}_{\rm 1}\over {\rm M}_{\odot}} \right ) {\rm M}_{\odot}\,{\rm yr}^{-1}.  
\end{equation}   
For the IMBH masses we consider, this is in excess of 10$^{-7} {\rm M}_{\odot}$ yr$^{-1}$, corresponding to L$_{\rm X}\sim{\rm few}\times 10^{40}$ erg s$^{-1}$.  Transient IMBH ULX sources are more plausible, because they confine the most extreme rates of mass transfer to short outburst periods.  Admittedly our current physical understanding of transient behavior is not complete.  Nevertheless, it is generally accepted that an accreting binary becomes a transient when the mass transfer rate driven by the donor is lower than a certain critical value and a thermal disk instability develops.  In what follows we adopt the critical mass transfer rate for transient behavior of an X-ray irradiated disk given by \citet{dub99}:
\begin{eqnarray}\label{eqn_mdotcrit}
%\small{ \dot M^{\rm irr}_{\rm crit}  \simeq  10^{-5} \left( {M_{\rm 1}\over M_{\odot}} \right )^{0.5} \left( {M_{\rm 2}\over M_{\odot}} \right )^{-0.2} \left( {P\over 1 {\rm yr}} \right )^{1.4}  M_{\odot}\,{\rm yr^{-1}} }
%\dot M^{\rm irr}_{\rm crit} \simeq 2.0 \times 10^{15} \left( {M_1 \over M_{\odot}} \right)^{0.5} \left( {M_2 \over M_{\odot}} \right)^{-0.2} P^{1.4}_{\rm hr} \times \left( {C \over 5\times10^{-4}} \right)^{-0.5} {\rm g s^{-1}}
\dot {\rm M}^{\rm irr}_{\rm crit} \simeq 1.5\times10^{15} \left( {{\rm M}_1 \over {\rm M}_{\odot}} \right)^{-0.4} \left( {R \over 10^{10} {\rm cm}} \right)^{2.1} \nonumber \\
\times \left( {C \over 5\times10^{-4}} \right)^{-0.5} {\rm g \; s^{-1}}
\end{eqnarray}
for donor mass M$_{\rm 2}$ and disk radius {\it R.  \it C} is assumed to be $5\times10^{-4}$ for most donors.  It can also be shown that the critical $\dot {\rm M}$ for transience corresponds to a minimum black hole mass that depends on companion mass and orbital period \citep{king96}.  As \citet{kal04} have shown, this minimum BH mass is high for main sequence (MS) donor stars ($10^{3} {\rm M}_{\odot}$ on average for a $10 {\rm M}_{\odot}$ donor).  Red giant (RG) donors can more easily form transient systems with IMBHs.  One should note that, due to the minimum mass limit, a stellar-mass BH binary is unlikely to be transient, so any transient ULX observed is a strong IMBH candidate. 

%\subsection{Methods of IMBH Formation}
In our simulations, we consider IMBHs that have formed through ``runaway collisions'' of MS stars.  This is a channel for IMBH formation in which a series of rapid mergers occur as the cluster begins core collapse, causing growth of a central massive object that can collapse to a black hole \citep{pz99,pm02,gfr04,frb05,fgr05}.  It has been suggested that an IMBH formed though runaway growth actually prevents core collapse and causes the core to reexpand \citep{bme04}.  In the dense clusters we are considering, the timescale for IMBH formation through runaway mergers is estimated to be $\la 3$ Myr \citep{pm02, gfr04}.  Because the core collapse time, t$_{\rm cc}$, is expected to obey the relation t$_{\rm cc} \sim 0.15$t$_{\rm rc}(0)$, where t$_{\rm rc}(0)$ is the initial core relaxation time, this constrains t$_{\rm rc}$ to be $\la 30$ Myr \citep{gfr04}.

Given that a plausible IMBH formation method exists, the question most pertinent to understanding ULXs is whether an IMBH, once formed, will gain close stellar companions that can sustain mass transfer at ULX X-ray luminosities.  No prior studies have attempted to investigate numerically with both dynamics and binary evolution whether binary formation and mass transfer can occur with IMBHs in a dense cluster core (for an analytical exploration of tidal capture and its consequences, see Hopman, Portegies Zwart, \& Alexander 2004).  Here, we take the first step in studying IMBH binary interactions as a ULX possibility, using detailed numerical simulations that combine dynamical interactions with full binary stellar evolution.  In \S\,2 we outline the details of our simulations.  The results for our standard cluster model (details below) are presented in \S\,3.  The variations of these results for other cluster models are described in \S\,4.  In \S\,5, we discuss our results, estimating a lower limit on the incidence of IMBH mass transfer, and we outline our goals for future simulations.

\section{Cluster Simulations with Central IMBH}

The goal of our simulations is to examine the ability of IMBHs in young, dense stellar clusters to form and maintain close binary systems with companion stars.  We employ a numerical code developed by Ivanova et al. (2005, where a detailed description of the method can be found). This adopted hybrid code incorporates: (i) a stellar population synthesis code to evolve single stars and binary systems ({\em StarTrack}; Belczynski, Kalogera, \& Bulik 2002; Belczynski et al.\ 2005 to be submitted); (ii) a semi-analytical prescription for 2-, 3-, and 4-body stellar collisions, disruptions, exchanges, and tidal captures based on previously derived cross sections; and (iii) a numerical toolkit for direct N-body integration of 3- and 4-body systems ({\em Fewbody}; Fregeau et al.\ 2004).  This combination provides us with a unique numerical tool that lends itself to the study of binary interactions in cluster cores.  At the young cluster ages we consider, large-scale cluster evolution has little effect, so certain simplifications can be made.  We employ a ``fixed background'' model, which holds the core stellar number density n$_{\rm c}$ and velocity dispersion $\sigma$ constant (e.g., Hut, McMillan, \& Romani 1992; Di Stefano \& Rappaport 1994; Sigurdsson \& Phinney 1995; Portegies Zwart et al. 1997a; Portegies Zwart et al. 1997b; Rasio et al. 2000).  The cluster density profile is assigned by dividing the cluster into two regions: a dense core and an outer halo.  This model is qualitatively based on mass segregation.  Note that the core is expected to expand on a very short timescale immediately after the IMBH is formed, smoothing out the cusp and supporting the assumption of a constant density core \citep{bme04}.

In our models, an IMBH is added to the core at 1 Myr, in accordance with estimated IMBH formation times of 1-3 Myr in clusters \citep{pm02}.  To form an IMBH through runaway collisions, the initial core relaxation time, ${\rm t_{rc}}$, must be $\la 25-30$ Myr \citep{gfr04}.  This time is given by:
\begin{equation}
{\rm t_{rc}} = { \sigma^3_{\rm 3D} \over 4.88 \pi G^2 {\rm ln}(\gamma {\rm N}) {\rm n_c} \langle m \rangle^2 }
\end{equation}
\citep{spitz87}, where $\sigma_{\rm 3D}$ is the three-dimensional velocity dispersion, n$_{\rm c}$ is the core number density, N is the total number of stars, $\langle m \rangle$ is the average stellar mass, and  $\gamma$, the coefficient in the Coulomb logarithm $\Lambda = \gamma {\rm N}$, is assumed to be 0.01 (cf. G\"urkan, Freitag, \& Rasio 2004).  For our standard model (parameters are described below), ${\rm t_{rc}} \simeq$ 29 Myr, in good agreement with the constraint for runaway collisions.

We evolve the clusters to 100\,Myr, as we consider only young clusters like those with which ULXs are most often associated.  We can then neglect long-term contributions of the outer halo to cluster evolution and synthesize the entire stellar population of N = $2.7\times10^{4}$ in the core.  Mass segregation within the core is incorporated through our choice of the initial mass function (IMF) used for the stellar population in the core.  

An initial binary fraction of 100\% is assumed, which provides the maximum possible concentration of binaries to interact with and become companions of the IMBH.  Observations indicate that young clusters are expected to have very high binary fractions \citep{lle99, apai04, dd04}.  This assumption is also consistent with the suggestion that cluster binary populations are depleted over time by interactions and stellar evolution \citep{mik83, hill84, bsd96, fre04, iva05}.  Initial conditions for binary orbital parameters are assigned as outlined in \S\,3.1 of \citet{iva05}, with slight modifications:

\begin{itemize}

\item The binary mass ratio $q = m_2/m_1$ is assigned a uniform distribution $0 < q < 1$, where M$_1$ and M$_2$ are the binary primary and secondary, respectively.  
\item A uniform logarithmic distribution is used for the binary period, P, ranging from $0.1 - 10^{7}$ d.  
\item The thermal distribution is used for the binary eccentricities $e$ with probability density $\rho(e) = 2e$. 
\item Binary systems are rejected from the initial distribution if one of the binary components enters Roche lobe overflow (RLOF) at pericenter.  
\item We use the IMF of \citet{k02} for primary stars between 0.8 and 100 M$_{\odot}$ in most simulations.  
\item Secondary stars are restricted in a mass range between 0.05 and 100\,M$_{\odot}$.

\end{itemize}

\begin{deluxetable}{llr}
\tablecaption{Cluster Models\label{cm}}
\tablewidth{0pc}
\tablecolumns{3}
\tablehead{
 \colhead{Model}
& \colhead{IMF}
& \colhead{n$_{\rm c}$ [pc$^{-3}$]}
}
\startdata
 A & Kroupa & $1.33\times10^{5}$ \\
 B & Flattened & $1.33\times10^{5}$ \\
 C & Kroupa & $1.33\times10^{4}$ \\
\enddata
\tablecomments{The IMF in models B is a broken power-law mass function $\xi({\rm M}) \sim {\rm M}^{\alpha}$, with $\alpha$ = -1.25 for $0.8 < {\rm M/M}_{\odot} < 5$ and $\alpha$ = -1.5 for ${\rm M/M}_{\odot} > 5$.  Models A \& B were tested with and without 3BBF.  Model A is also tested with a range of IMBH masses from 50 to 500 M$_{\odot}$.} 
\end{deluxetable}

Our standard cluster model (model A, see Table~\ref{cm}) has a 1-D stellar velocity dispersion $\sigma = 10$ km s$^{-1}$ and a core stellar number density n$_{\rm c} = 1.33\times10^5$ pc$^{-3}$.  The latter corresponds to a luminosity density $\geq 10^{7} {\rm L}_{\odot}$ pc$^{-3}$ for a cluster age $\leq 10^{8}$ yr.  The total cluster mass is $\simeq 5\times10^4 {\rm M}_{\odot}$.  We use 100 M$_{\odot}$ as our standard IMBH mass, but we explore a range of IMBH masses from $50 - 500 {\rm M}_{\odot}$ (see \S\,4.1).  Note that each type of simulation must be repeated many times to obtain good statistics.

We also consider two variations of our standard cluster model (see Table~\ref{cm}): with models B we explore the effect of a flatter IMF, and with models C we examine results for less dense clusters. Detailed simulations of clusters that form an IMBH just before core collapse have shown that rapid mass segregation preceding IMBH formation creates a mass function flatter than Kroupa's for stars $> 5-10\,{\rm M}_{\odot}$ (G\"urkan, Freitag, \& Rasio 2004; Freitag, G\"urkan, \& Rasio 2005; Freitag 2005, personal communication). These results have prompted us to consider the effects of mass segregation on the formation of IMBH binaries. We have also examined whether any Brownian-like motion of the IMBH would cause it to interact with stellar populations that have different properties than those typical of the compact cluster cores. Assuming energy equipartition exists between the IMBH and the cluster, the IMBH in a constant-density core will experience radial oscillations of amplitude \citep{bw76}
\begin{equation}
R \simeq {R_{\rm core} \over \sqrt{3}} \left( {\rm M}_{\rm IMBH} \over {\rm M}_{\rm tot} \right)^{1/2}, 
\end{equation}
where $R_{\rm core}$ is the core radius, M$_{\rm IMBH}$ is the IMBH mass, and M$_{\rm tot}$ is the total mass of the rest of the cluster.  For a 100 M$_{\odot}$ IMBH and total cluster mass $\simeq 5\times 10^4 {\rm M}_{\odot}$, we obtain an oscillation amplitude that is safely small: $R/R_{\rm core} \simeq 2.5\%$.  

We do not normally allow binary formation through three-body interactions in any of our simulations, as the initial binary fraction is 100\%, but we do test the effect of 3-body binary formation (3BBF) on models A and B. The details of our treatment of 3BBF are given in \citet{iva05}. 

A large fraction of IMBH binaries acquire at least one, and often a series of tertiary objects; however, no population synthesis code currently exists that is capable of modeling full stellar evolution for triple systems. For triple systems that are considered stable after the detailed direct N-body modeling with {\em Fewbody} (see stability criterion in Mardling \& Aarseth 2001), we implement a ``breakage'' prescription that is physically self-consistent. The energy required to eject the outer star is obtained by shrinking the inner binary orbit.  The outer star is released unless the inner binary merges during shrinkage; in this case the inner system is allowed to merge and the outer companion is kept at its new, wider orbit to form the final binary system.  The condition for merger is that the stars in the inner binary are in physical contact at pericenter or, in the case of an IMBH binary, the secondary star enters the black hole's tidal radius 
\begin{equation}
R_{\rm tidal} \approx R_2 \left ({{\rm M}_1\over {\rm M}_2}\right )^{1/3}.
\end{equation}
We recognize that this artificial treatment can affect the evolution of IMBH binaries, for example, by preventing eccentricity boosts via the Kozai mechanism \citep{k62}.  For this reason we have undertaken a careful analysis of triple breaking and its effect on our results (see \S\,3.5).  It should also be noted that in some cases with a RG companion, release of the outer companion induces a common-envelope phase in the resulting binary.

\section{Results for Standard Model}

In what follows we present the results of our standard, 100 M$_{\odot}$ IMBH cluster simulations (model A in Table 1) in the context of the formation of IMBH binaries that experience mass transfer and may become observable as ULXs. In \S\,3.1 we discuss the characteristics of IMBH companions, including their orbital parameters, evolutionary types, and mass transfer phases.  In \S\,3.2 we present an example of a single simulation and more details on the history of the IMBH.  In \S\,3.3 we examine the X-ray luminosities produced by mass transfer and determine the time fraction over which the IMBH could be observable as a ULX.  In \S\,3.4 we comment on the statistical interpretation of our results.  Finally, in \S\,3.5 we analyze triple stellar systems and their possible effect on mass-transferring binaries.

\begin{figure}
\epsscale{1.2}
\resizebox{\hsize}{!}{\includegraphics{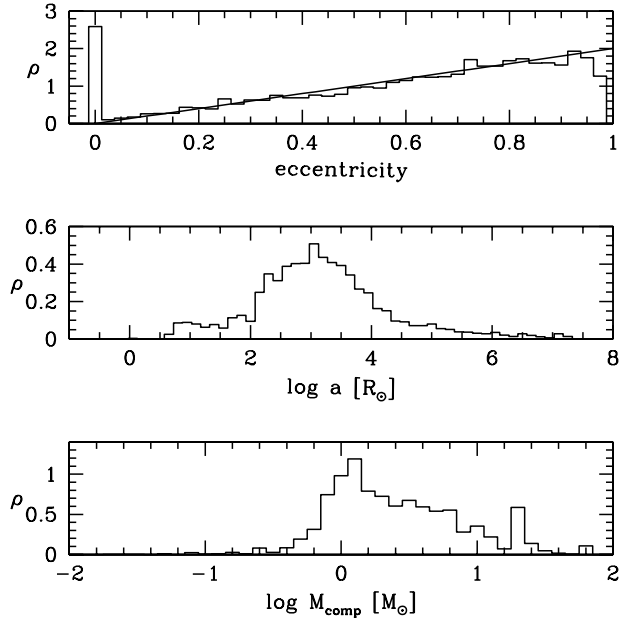}}
%\plotone{figure1.eps}
\caption{Probability densities of IMBH binary properties throughout the simulations: companion mass, orbital separation and eccentricity.  The thermal distribution, $\rho(e)=2e$, is shown on the eccentricity plot. Results from 167 simulations of model A with a 100 M$_{\odot}$ IMBH are shown.\label{dist}}
\end{figure}

\subsection{Statistical Results on IMBH Companions}

\begin{figure}
\epsscale{1.2}
%\epsscale{0.5}
\resizebox{\hsize}{!}{\includegraphics{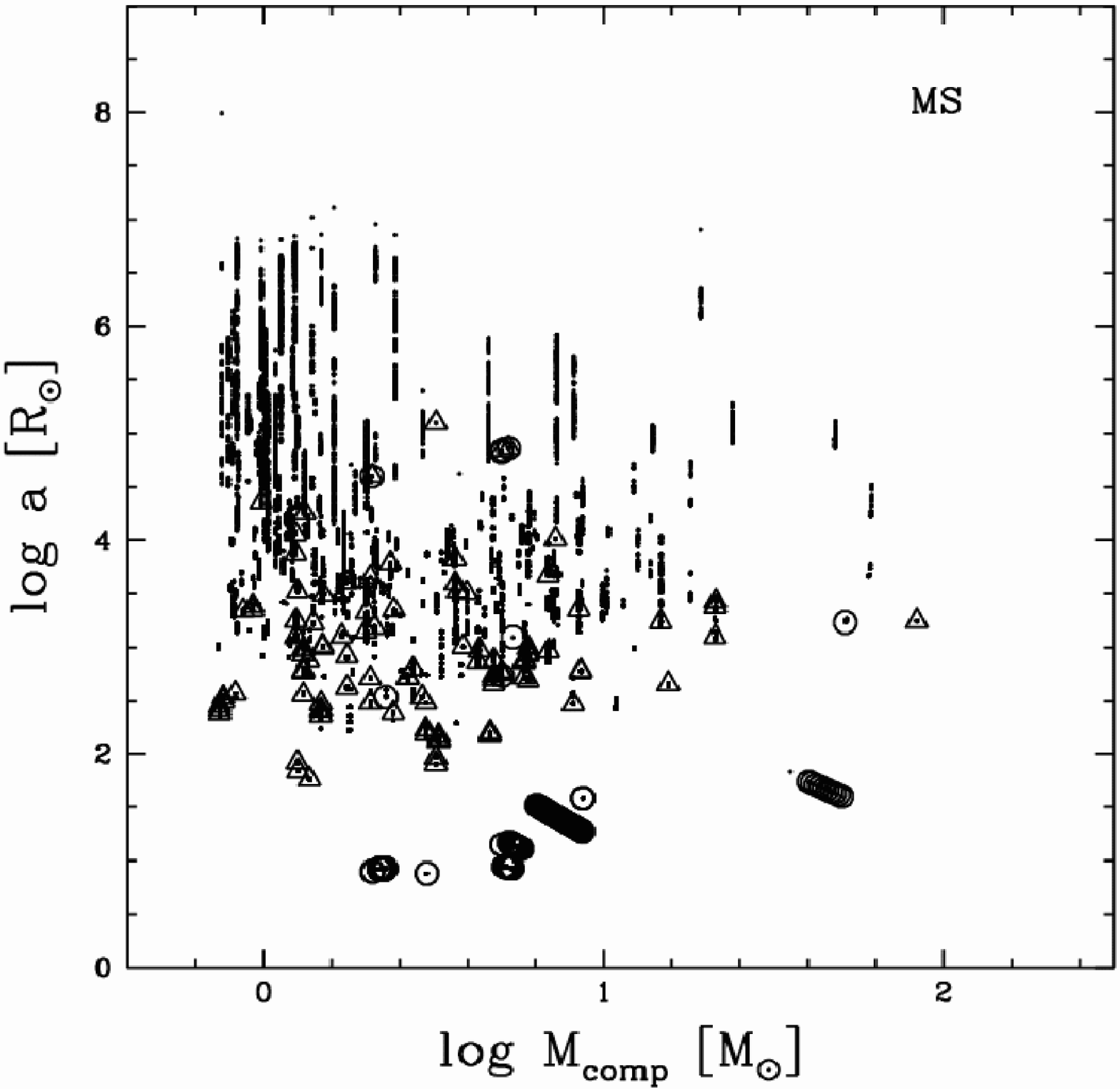}}
%\epsscale{0.5}
\resizebox{\hsize}{!}{\includegraphics{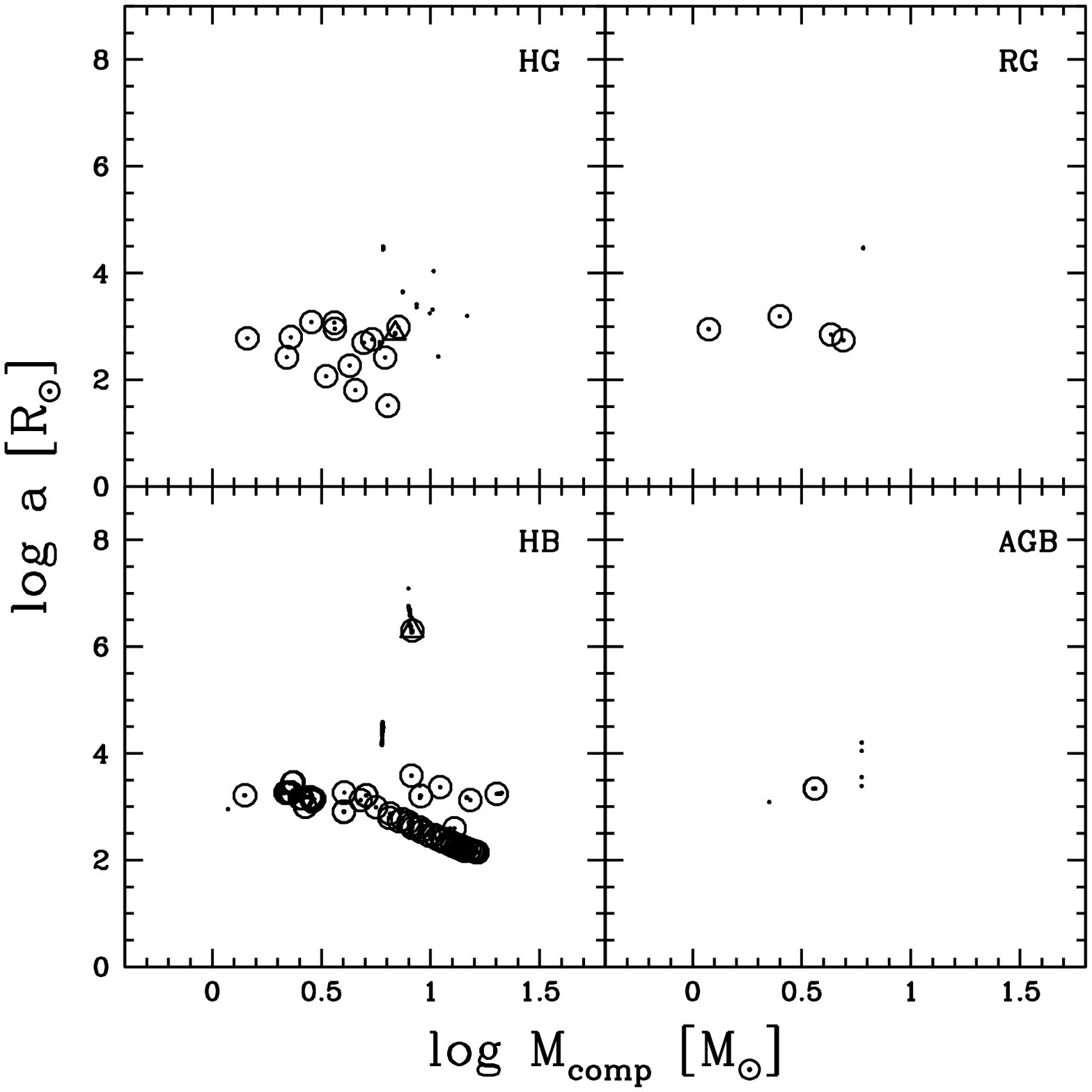}}
%\plottwo{figure2a.eps}{figure2b.eps}
%\epsscale{0.6}
%\plotone{figure2a.eps}
%\plotone{figure2b.eps}
\caption{Companion mass vs. orbital separation for 42 model A simulations with a 100 M$_{\odot}$ IMBH, presented separately for the following companion evolutionary types: MS, Hertzsprung gap (HG), first giant branch (RG), horizontal branch (HB), and asymptotic giant branch (AGB).  Fewer He and WD companions undergo mass transfer, and they are not included here.  Dots denote simulation timesteps when the IMBH has a binary companion.  Open circles denote a mass-transferring companion (through RLOF).  Open triangles denote binary companions associated with a triple system that was broken.\label{evtype}}
%\epsscale{}
\end{figure}

The distributions of three orbital parameters: companion mass, orbital separation, and eccentricity, are shown in Fig.~\ref{dist} for all $100 {\rm M}_{\odot}$ IMBH simulations for cluster model A.  The prevalence of low-mass companions ($\sim1 {\rm M}_{\odot}$) in the core is consistent with the model IMF that is dominated by such low-mass stars.  Orbital separations of the dynamically-formed IMBH binaries range widely from $\sim1-10^{7} R_{\odot}$.  The lower limit corresponds to the Roche lobe radius of the IMBH companion.  On the other end, the soft boundary for binaries is defined as the ``escape radius''
\begin{equation}
{\rm a}_{\rm soft} = {{\rm G}\, {\rm M}_{\rm IMBH}\over \sigma^2} \simeq 10^5 R_{\odot}.
\end{equation}
In the simulations, a star is treated as a companion if its orbital separation is less than $100 {\rm a}_{\rm soft}$, corresponding to the maximum orbital separation, $10^7 R_{\odot}$.  Fig.~\ref{dist} illustrates that the majority of IMBH companions have orbital separations between 100 and $10^{4} R_{\odot}$.  Eccentricities follow the initial thermal distribution, which favors higher values (Fig.~\ref{dist}).  These high eccentricities tend to increase the occurrence of mass-transfer events in which the companion overflows its Roche lobe at pericenter.

We find that the IMBH has a companion for $\simeq$60\% of the total simulation time, and it has a mass-transferring companion for only $\simeq$3\% of the total evolution time (Table~\ref{msrg}).  Fig.~\ref{evtype} shows companion masses and orbital characteristics for different evolutionary types.  The IMBH companions are overwhelmingly MS stars due to the much longer lifetime of this phase and their dominance by number.  About 90\% of unique 100 M$_{\odot}$ companions are MS stars for at least part of their binary lifetime; the IMBH has an average of $\sim$6 MS companions per simulation and only one post-MS companion for every two simulations [``post-MS'' will henceforth refer to all post-main-sequence evolutionary types capable of driving mass transfer - i.e., neutron stars (NSs) and BHs are excluded].  In contrast, post-MS companions are more likely candidates for RLOF than are MS companions.  The percentage of post-MS companions that undergo mass transfer in our simulations is much higher than the percentage of mass-transferring MS companions (see Table~\ref{msrg}).   As noted above, post-MS companions are also more likely to enter transient mass transfer phases, which are more conducive to ULX formation.  However, because the MS lifetime is about 10 times longer, mass transfer phases of MS companions are typically of longer duration, and the {\it numbers} of MS and post-MS mass-transferring companions per simulation are comparable.  (Table~\ref{msrg}).

\begin{deluxetable}{llcrr}
\tablecaption{IMBH binary companions \label{msrg}}
%\tabletypesize{\scriptsize}
\tablewidth{0pc}
\tablecolumns{5}
\tablehead{
\colhead{}
&\colhead{}
&\colhead{M$_{\rm IMBH}$ = 100 M$_{\odot}$}
&\colhead{}
&\colhead{}
}
\startdata
%\multicolumn{2}{c}{M$_{\rm IMBH}$ = 100 M$_{\odot}$} \\
 & MC Runs & & 167 & \\
& $\langle {\rm N_{comp}} \rangle$ & & 6.49 & \\ & & & & \\
& $\langle {\rm N_{MS}} \rangle$ & & 5.90 & \\ & & & & \\
& $\langle {\rm N_{postMS}} \rangle$ & & 0.47 & \\ & & & & \\
& $\langle {\rm N^{MT}_{MS}} \rangle$ & & 0.26 & \\ & & & & \\
& $\langle {\rm N^{MT}_{postMS}} \rangle$ & & 0.23 & \\ & & & & \\
& \% MS$_{\rm MT}$ & & 4.36 & \\ & & & & \\
& \% postMS$_{\rm MT}$ & & 50.00 & \\ & & & & \\
& \% t$_{\rm comp}$ & & 57.63 & \\ & & & & \\
& \% t$_{\rm MS}$ & & 40.87 & \\ & & & & \\
& \% t$_{\rm postMS}$ & & 4.44 & \\ & & & & \\
& \% t$^{\rm MT}_{\rm MS}$ & & 2.81 & \\ & & & & \\
& \% t$^{\rm MT}_{\rm postMS}$ & & 0.11 & \\
\enddata
\tablecomments{All simulations adopt cluster model A.  MC Runs: number of distinct Monte Carlo simulations evolved to $10^8$\,yr.  $\langle {\rm N_{comp}} \rangle$: average number of companions (of any type) per run.  $\langle {\rm N_{MS}} \rangle$ \& $\langle {\rm N_{postMS}} \rangle$: average number of companions of each type per run.  Most postMS companions are captured while on the MS and do not correspond to different stellar companions.   $\langle {\rm N^{MT}_{MS}} \rangle$ \&  $\langle {\rm N^{MT}_{postMS}} \rangle$: average number of MS and post-MS companions that undergo mass transfer per run.  \% MS$_{\rm MT}$ \& \% postMS$_{\rm MT}$: number \% of MS and post-MS companions that undergo mass transfer.  \% t$_{\rm comp}$: average \% of total simulation time spent with a companion of any type.  \% t$_{\rm  MS}$ and \% t$_{\rm  postMS}$: average \% of time with MS and post-MS companions.  \% t$^{\rm MT}_{\rm MS}$ \& \% t$^{\rm MT}_{\rm postMS}$: average \% of time spent in MS and post-MS mass transfer phases.}

\end{deluxetable}

\subsection{Single Run Example}

\begin{figure}
\epsscale{1.2}
\resizebox{\hsize}{!}{\includegraphics{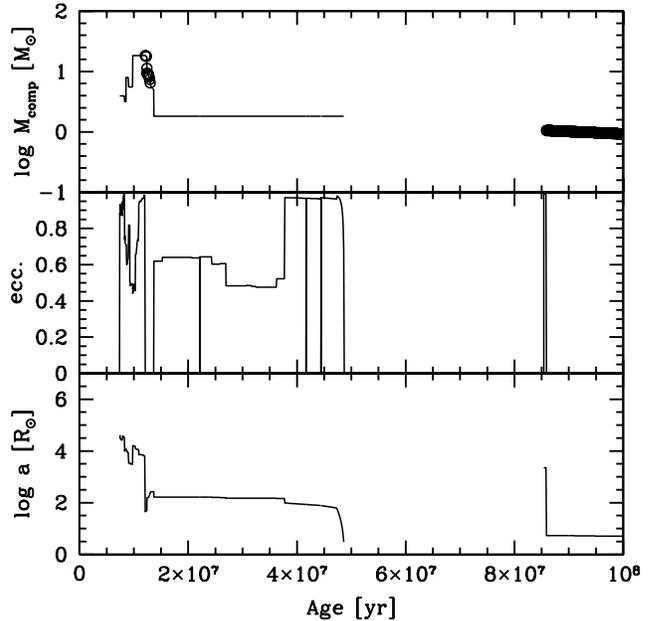}}
%\plotone{figure3.eps}
\caption{Evolutionary history of a single 100 M$_{\odot}$ IMBH simulation.  Top window: IMBH companion mass M$_2$; middle window: eccentricity; bottom window: orbital separation.  Circles denote timesteps with mass transfer; each non-mass-transferring companion appears at a different mass value.  It is evident that the IMBH changes muliple binary companions during a single cluster simulation. Gaps in the top and bottom plots indicate times with no IMBH companion. \label{evhist}}
\end{figure}

\begin{figure}
\epsscale{1.2}
\resizebox{\hsize}{!}{\includegraphics{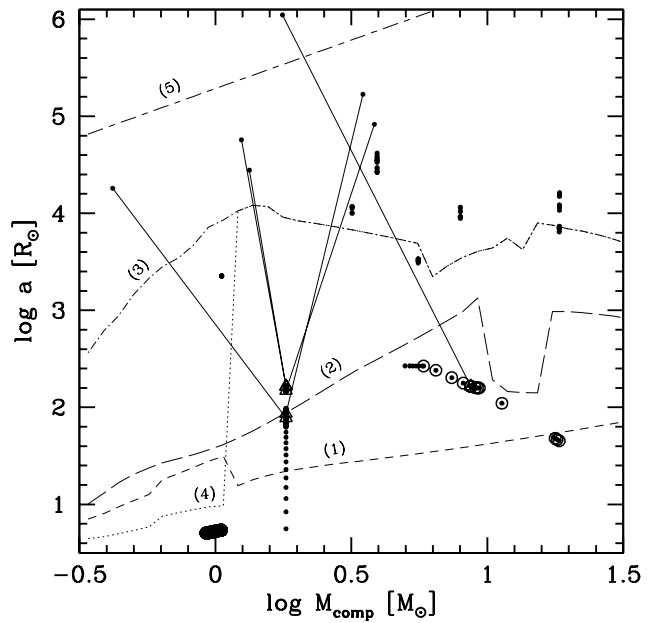}}
%\plotone{figure4.eps}
\caption{IMBH companion mass (M$_2$) vs. orbital separation (a) for the same simulation shown in Fig.\,1.  Dots denote timesteps with an IMBH companion; open circles denote mass-transferring companions.  Triangles mark inner-binary orbital parameters of triple systems and solid lines connect to the outer orbital parameters.   The maximum companion orbit radii for RLOF are shown at: the end of MS (line 1), the start of the RG branch (line 2), the maximum possible stellar radius (line 3), and the maximum stellar radius for clusters younger than $10^8$ yr (line 4).  Line 5 marks the soft/hard binary boundary. \label{1win}}
\end{figure}

The binary companion history of a single 100 M$_{\odot}$ IMBH simulation, using model A, is shown in Figs.~\ref{evhist} \&~\ref{1win}.  This simulation is atypical in that it has multiple mass transfer events, but the behavior of these events is characteristic of mass transfer seen in other simulations.   Fig.~\ref{evhist} shows the IMBH evolution over time.  The black hole has a companion for $\sim60\%$ of its total evolution time, which is about the average we calculate for all 100 M$_{\odot}$ simulations (Table~\ref{msrg}).  Note the often high eccentricities of non-mass-transferring companions, which is also typical of our simulations (Fig.~\ref{dist}).  In the lower window of Fig.~\ref{evhist}, one can see the hardening of several companion orbits before they are disrupted by another interaction.  The long mass transfer phase at the end of the simulation characterizes stable, MS mass transfer.  The first companion that undergoes RLOF (at about 10$^7$ yr) is the most interesting.  Mass transfer begins on the MS and continues through the HG until the donor becomes a He HG star.  After this, the star explodes and leaves a NS orbiting the IMBH.  The companion's circularized orbit is kicked by the supernova to an eccentricity of about 0.6.  As Fig.~\ref{evhist} shows, the NS is later forced into an even higher-$e$ orbit by an encounter with an outer companion shown in Fig.~\ref{1win}.  This causes an eventual merger with the IMBH.  The outer-companion swapping that occurs with the NS-IMBH binary is frequently-observed behavior for a close, stable binary; more details on the behavior of triple systems are presented in \S\,3.5.

\subsection{X-ray Luminosities}

We examine the range of X-ray luminosities produced by the various mass-transfer phases that occur in our standard model simulations. 
The binary evolution component of the simulations allows us to calculate the mass transfer rate $\dot {\rm M}$ driven by RLOF donors to the IMBH.  We compare this rate to the critical rate $\dot {\rm M}_{\rm crit}$ for transient behavior (Eqn.~\ref{eqn_mdotcrit}) to determine whether the IMBH XRB is persistent or transient. If persistent, the transfer rate from the donor is converted to an expected X-ray luminosity; if transient, we assume that the XRB emits at its Eddington luminosity during disk outbursts \citep{kal04,ik05}. In model A clusters with a 100 M$_{\odot}$ IMBH, the total mass-transfer time fraction is about 3\%. Of this, about 10\% is spent under transient conditions, i.e., $\dot {\rm M} < \dot {\rm M}_{\rm crit}$.  However, only a very small fraction of the transient mass transfer time, comparable to the transient duty cycle, will be spent in outburst. Observations of Galactic transients imply that this duty cycle is of the order of a few percent \citep{mr05}. Such short duty cycles correspond to very low time fractions of the $10^8$\,yr cluster age that could be associated with ULX emission: $< 0.03\%$. The probability of detecting one of these transient sources in outburst is unfortunately vanishingly low.  
The other 90\% of mass transfer time in our simulations is persistent and a ULX will result, by definition, if the X-ray luminosity exceeds $10^{39}$ erg s$^{-1}$. We find that a 100\,M$_{\odot}$ IMBH becomes a persistent ULX for only $\sim$2\% of total mass transfer time in model A clusters; this corresponds to $\simeq 0.06$\% of the total age of the young clusters we consider here.  We conclude that either persistent or transient IMBH ULXs would be observable for only a tiny fraction ($< 0.1$\%) of the cluster lifetime.

\subsection{Averages vs. Medians}

\begin{figure}
\epsscale{1.2}
\resizebox{\hsize}{!}{\includegraphics{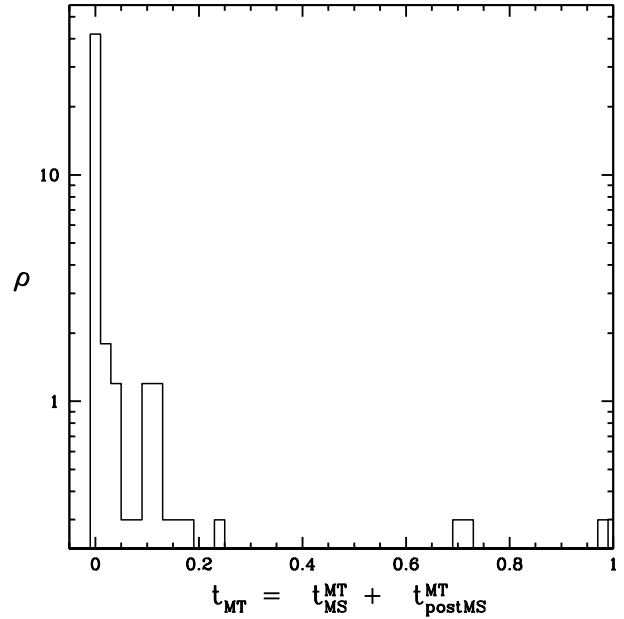}}
%\plotone{figure5.eps}
\caption{Distribution of mass transfer time fraction for 167 standard model simulations with a 100 M$_{\odot}$ IMBH.\label{MTfrac}}
\end{figure}

It should be emphasized that the mass transfer time fractions we present are averaged over all simulations, and that individual simulations are given equal weight, but their results do vary over a wide range.  The majority of our simulations have {\em no} IMBH mass-transfer events at all; it turns out that the {\em median} mass-transfer time fraction for each cluster model is always at or very near zero (Fig.~\ref{MTfrac}).  At the other extreme, IMBHs in a few model A simulations acquire very stable mass-transferring companions and spend more than half of their total evolution time in mass transfer phases. It is clear that the averages reported are affected by the presence of these rare outliers, as one can see by comparing the entire model A data set (Table~\ref{msrg}, t$^{\rm MT}_{\rm MS}$ and t$^{\rm MT}_{\rm postMS}$) to the smaller set for which triple system data was obtained, which does not include the outliers (Table~\ref{trp}, t$_{\rm MT}$).  We would like to note that for the majority of our simulations the predictions for the IMBH binaries leading to mass transfer with X-ray luminosities in the ULX regime could be even lower than we report based on the average quantities and results.

\subsection{Triple Systems}

The significant number of triple systems that form and are artificially (although physically self-consistently, i.e., conserving energy) disrupted in our simulations warrant a separate analysis of their possible effect on mass transfer.  The NS merger event shown in the example simulation (Figs.~\ref{evhist} and~\ref{1win}), which resulted from an eccentricity boost by a third object, indicates that three-body interactions can play an important role even when triples are artificially broken.  By analyzing the initial orbital characteristics of triples and their coincidence with mass-transfer phases, we develop a sense of whether our treatment of triples is, in fact, a reasonable approximation.

Almost all of the triples are hard in the outer orbit (see Fig.~\ref{1win}), but depending on orbital parameters, the inner binaries may not be influenced by the outer companion.  We quantify the degree of possible companion interaction using the ratio of the outer orbit's pericenter to the inner orbit's apocenter ($\wp_{\rm out}/ {\rm a}_{\rm in}$), that is, the closest possible approach between the two bodies.  The result is shown in Fig.~\ref{aratio} (upper left window) and in Table~\ref{trp}.  We assign $\wp_{\rm out}/ {\rm a}_{\rm in} \leq 3$ as the critical value of this ratio, below which it is likely that the outer companion will have a nonnegligible effect on the inner binary.  These ``close'' systems comprise less than 1/3 of all triples.   

Another diagnostic for triple systems is to calculate the percentage of mass transfer phases potentially affected.  We consider ``affected'' mass transfer as that which follows or coincides with triple formation.  More importantly, we also calculate the percentage of mass transfer phases associated with close triples.  While almost half of mass-transferring companions are affected, only $\sim10\%$ are affected by close triples (Table~\ref{trp}). If we instead consider the total mass transfer {\it time} that is similarly affected, this fraction drops to $\sim5\%$.

\begin{deluxetable}{llcrr}
\tablecaption{Triple systems and IMBH companions \label{trp}}
%\tabletypesize{\scriptsize}
\tablewidth{0pc}
\tablecolumns{5}
\tablehead{
\colhead{}
&\colhead{}
&\colhead{M$_{\rm IMBH}$ = 100 M$_{\odot}$}
&\colhead{}
&\colhead{}
}
\startdata
& MC Runs & & 109 & \\ & & & & \\
& $\langle {\rm N_{trip}} \rangle$ & & 3.46 & \\ & & & & \\
& $\langle {\rm N_{trip}/N_{bin}} \rangle$ & & 0.50 & \\ & & & & \\
& \% ${\rm trip_{close}}$ & & 28.90 & \\ & & & & \\
& \% MT$_{\rm aff}$ & & 47.72 & \\ & & & & \\
& \% MT$_{\rm close}$ & & 10.24 & \\ & & & & \\
& \% t$_{\rm MT}$ & & 1.60 & \\ & & & & \\
& \% t$^{\rm MT}_{\rm aff}$ & & 21.07 & \\ & & & & \\
& \% t$^{\rm MT}_{\rm close}$ & & 4.96 & \\
\enddata

\tablecomments{All simulations adopt cluster model A.  MC Runs: number of distinct Monte Carlo simulations for which triples data were obtained (note that Tables~\ref{trp} and~\ref{trpmass} are based on smaller sample sizes than are other data).  $\langle {\rm N_{trip}} \rangle$: average number of unique triple systems per run.  $\langle {\rm N_{trip}/N_{bin}} \rangle$: average ratio of distinct triple systems to distinct binary systems formed.  \% ${\rm trip_{close}}$: \% of triples that are ``close'' ($\wp_{\rm out}/ a_{\rm in} \leq 3$).   \% MT$_{\rm aff}$: \% of MT companions that are potentially affected by one or more tertiary companions gained during their lifetime.  \% MT$_{\rm close}$: same fraction for close triples.  \% MT t$_{\rm aff}$: percent of total MT time potentially affected by triples.  \% MT t$_{\rm close}$: same fraction for close triples.}
\end{deluxetable}

Apart from calculating the fractions of binaries and mass transfer phases that could be affected by our treatment of triples, we also attempt to quantify the possible effects of triple breaking on the mass transfer phases.  One way in which an outer companion may affect the inner binary is through the Kozai mechanism \citep{k62} that can increase the eccentricity of the inner binary and lead to mass transfer. We cannot exclude a priori that by breaking the triples we miss some mass transfer events involving the IMBH. We calculate the timescale for the inner binary to reach maximum eccentricity due to the Kozai mechanism using the initial orbital parameters of the triples \citep{inn97,mt98,wen03}:
\begin{equation}
{\tau}_{\rm koz} \simeq 0.16 f \left ( {\rm M}_{\rm i} \over {\rm M}_{\rm o}^2 \right )^{0.5} \left ( {\rm a}_{\rm o}^3 \over {\rm a}_{\rm i} \right )^{0.5} (1-e_{\rm o}^2)^{1.5} \, {\rm yr},  
\end{equation}
where $f \simeq 0.42 \, {\rm ln}(1/e_{\rm i0}) / ({\rm sin}^2(i_0)-0.4)^{0.5}$, $e_{\rm i0}$ and $i_0$ are the initial inner eccentricity and inclination, $e_{\rm o}$ is the outer eccentricity, M$_{\rm i}$ is the total mass of the inner binary, M$_{\rm o}$ is the outer companion mass, a$_{\rm i}$ and a$_{\rm o}$ are the inner and outer orbital separations, masses are in M$_{\odot}$, and distances are in AU.  If $\tau_{\rm koz}$  is sufficiently short, the outer companion left alone could cause the inner binary to undergo mass transfer at pericenter. However, this would be possible only if $\tau_{\rm koz}$ is shorter than the collision timescale, $\tau_{\rm coll}$, appropriate for the triple system:
%\begin{equation}
\begin{eqnarray}
%{\tau}_{\rm coll} = 1.7 \times 10^8 {\rm yr} \; \eta^2 k^{-2} {\rm n}_5^{-1} {\langle M\rangle^2\over M_{\rm i}^2 M_{\rm o}^2} \nonumber \\
%\times \left ( 1 + \eta {2\over k} { M_{\rm tot} + \langle M\rangle\over M_{\rm i} M_{\rm o} } \langle M\rangle \right )^{-1},
\tau_{\rm coll} \simeq 3.4\times10^{13} \times P^{-4/3}_{\rm d} {\rm M}^{-2/3}_{\rm tot} {\rm n}^{-1}_5 v^{-1}_{10} \nonumber \\
\times \left[ 1 + 913 {{\rm M}_{\rm tot} + \langle {\rm M} \rangle \over k P^{2/3}_{\rm d} {\rm M}^{1/3}_{\rm tot} v^2_{10}} \right]^{-1},
\end{eqnarray}
%\end{equation}
where $\langle {\rm M} \rangle$ is the average single star mass, n$_5$ is the cluster's stellar number density in units of 10$^5 {\rm pc}^{-3}$, $v_{10}=v_{\infty}/(10\,{\rm km\; s^{-1}})$, $v_{\infty}$ is the relative velocity at infinity, $P_{\rm d}$ is the binary period in days, and a strong encounter is defined by the closest approach d$_{\rm max}\leq ka$, with $k\simeq2$ \citep{iva05}.  We find that {\em only} $\sim11\%$ of all triples have $\tau_{\rm koz} < \tau_{\rm coll}$.  If these systems were not artificially disrupted, they could potentially induce mass transfer through Kozai resonance that is not seen in our simulations. However, we also note that disrupting a triple changes the properties of the inner binary, and this can also lead to mass transfer. We find that about $7\%$ of all tertiary companions are broken before or during a mass tranfer phase in inner binaries; this is the maximum fraction of broken triples in our set of standard simulations that could artificially create mass transfer by shrinking the inner binary.  Furthermore, we have examined the simulation data and found that there is essentially no overlap between disrupted triples that precede mass transfer and triples with $\tau_{\rm koz} < \tau_{\rm coll}$. Because the two populations are comparable in size (7\% and 11\%, respectively), we conclude that these possible effects on mass transfer events more or less mutually cancel.

It should be noted that a small fraction ($\sim4-8\%$) of triple systems form when a tight binary is captured by the IMBH; in this case, the IMBH is recorded as the outer, ``ejected'' companion and binary evolution is subsequently followed for only the two closely orbiting companions.  The data from this small number of unusual disrupted triples are not included in our results and has a negligible effect on the overall IMBH binary evolution.

\section{Parameter Study of Clusters with Central IMBH}

\subsection{IMBH Mass}

\begin{deluxetable}{lrrrrr}
\tablecaption{IMBH binary companions \label{mtmass}}
\tablewidth{0pc}
\tablecolumns{6}
\tablehead{
\colhead{ M$_{\rm IMBH}$ [M$_{\odot}$]}
& \colhead{50}
& \colhead{100}
& \colhead{150}
& \colhead{200}
& \colhead{500}
}
\startdata

MC Runs & 29 & 167 & 79 & 33 & 44 \\ & & & & & \\
$\langle {\rm N_{comp}} \rangle$ & 2.48 & 6.49 & 10.52 & 12.03 & 22.66 \\ & & & & & \\
$\langle {\rm N_{MS}} \rangle$ & 2.07 & 5.90 & 9.57 & 11.39 & 20.98 \\ & & & & & \\
$\langle {\rm N_{postMS}} \rangle$ & 0.34 & 0.47 & 0.84 & 0.55 & 0.98 \\ & & & & & \\
$\langle {\rm N^{MT}_{MS}} \rangle$ & 0.10 & 0.26 & 0.46 & 0.48 & 0.91 \\  & & & & & \\
$\langle {\rm N^{MT}_{postMS}} \rangle$ & 0.14 & 0.23 & 0.38 & 0.21 & 0.32 \\ & & & & & \\
\% MS$_{\rm MT}$ & 5.00 & 4.36 & 4.76 & 4.25 & 4.33 \\ & & & & & \\
\% postMS$_{\rm MT}$ & 40.00 & 50.00 & 45.45 & 38.89 & 32.56 \\ & & & & & \\
\% t$_{\rm comp}$ & 24.98 & 57.63 & 70.04 & 76.51 & 89.69 \\ & & & & & \\
\% t$_{\rm MS}$ & 18.76 & 40.87 & 53.69 & 60.89 & 56.95 \\ & & & & & \\
\% t$_{\rm postMS}$ & 3.59 & 4.44 & 4.05 & 1.77 & 2.36 \\ & & & & & \\
\% t$^{\rm MT}_{\rm MS}$ & 0.21 & 2.81 & 0.76 & 1.71 & 2.43 \\ & & & & & \\
\% t$^{\rm MT}_{\rm postMS}$ & 0.02 & 0.11 & 0.16 & 0.05 & 0.10 \\

\enddata

\tablecomments{All simulations adopt cluster model A.  Abbreviations given in Table~\ref{msrg}.}

\end{deluxetable}

\begin{figure}
\epsscale{1.2}
\resizebox{\hsize}{!}{\includegraphics{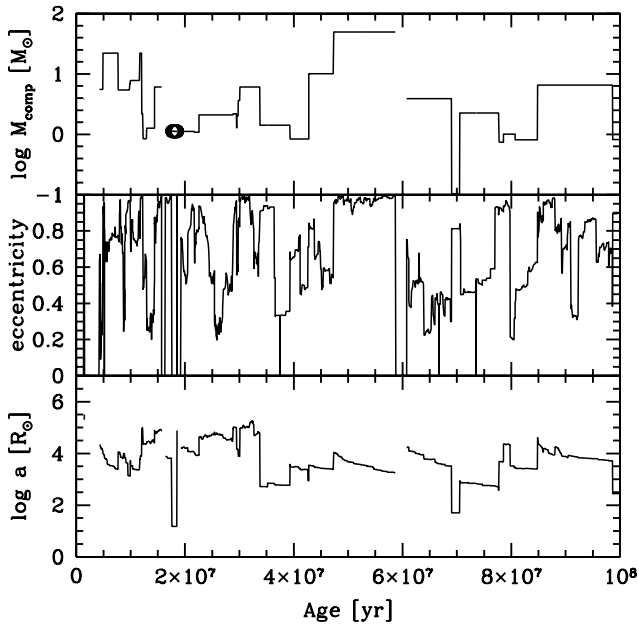}}
%\plotone{figure6.eps}
\caption{Increase in companion acquiring rate for a typical 500\,M$_{\odot}$ simulation over a 100 M$_{\odot}$ simulation (cf. Fig. 1) is shown; the IMBH has a companion for almost all of its lifetime.  Top window: companion mass; middle window: eccentricity; bottom window: orbital separation. \label{evhist500}}
\end{figure}

\begin{figure}
\epsscale{1.2}
\resizebox{\hsize}{!}{\includegraphics{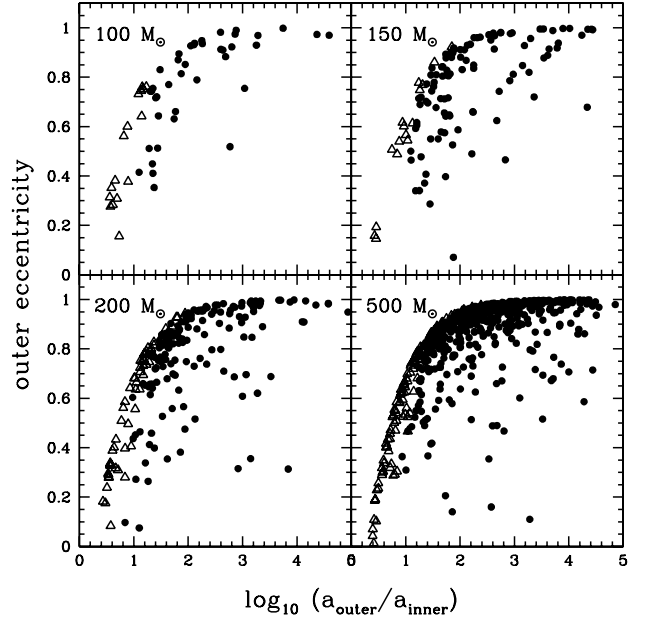}}
%\plotone{figure7.eps}
\caption{Triple systems formed for 16 simulations of 100, 150, 200, and 500 M$_{\odot}$ IMBHs.  The outer orbit's eccentricity is plotted vs. the ratio of orbital separations for each triple system.  Systems for which $\wp_{\rm out}/ a_{\rm in}$ is $\leq 3$ are denoted by open triangles; other triples are closed circles.\label{aratio}}
\end{figure}

Table~\ref{mtmass} outlines the evolutionary differences between IMBH masses in model A clusters.  The overall increase in companion capture with IMBH mass is apparent.  Note that the fractions of both MS and post-MS companions that undergo mass transfer are fairly consistent for all IMBH masses.  This underscores the greatly increased likelihood for a post-MS companion to undergo mass transfer compared to a MS companion, independent of external factors.  The time fractions spent with a companion of {\it any} type also increase monotonically with IMBH mass, but because BH companions are more prevalent at higher masses, this behavior is not as clearly seen for MS or post-MS companions.  The evolutionary history of a typical single 500 M$_{\odot}$ simulation is shown in Fig.~\ref{evhist500} to illustrate the large fraction of its lifetime spent with a companion.  

A clear result in Table~\ref{mtmass} is that 50 M$_{\odot}$ IMBHs have a very low rate of companion capture and mass transfer events.  This places a lower limit on the IMBH mass likely to form a luminous X-ray binary.  For higher IMBH masses, we emphasize as in \S\,3.4 that the average mass transfer time fractions are typically dominated by a few simulations with very long mass transfer phases.  Consequently, we do not observe any significant difference in $t^{\rm MT}_{\rm MS}$ or $t^{\rm MT}_{\rm postMS}$ between 100-500 M$_{\odot}$ IMBHs.  

As in 100 M$_{\odot}$ IMBH clusters, X-ray luminosities for other IMBH masses produce transient or persistent ULXs for insignificant fractions of the cluster lifetime, except possibly for the 500\,M$_{\odot}$. Clusters with 500 M$_{\odot}$ IMBHs do spend significantly more of their mass transfer time as ULX sources; these systems are transient for $\simeq$40\% of mass transfer time and persistent ULXs for $\simeq$12\% of mass transfer time.  Even so, these ULXs would be observable for only $\simeq0.35$\% of the total cluster lifetime.

The rate of triple system formation increases monotonically with IMBH mass (Fig.~\ref{aratio}, Table~\ref{trpmass}).  Clusters with a 50 M$_{\odot}$ IMBH have very few triple systems, as is expected based on their low rate of IMBH binary formation.  500 M$_{\odot}$ IMBHs have by far the highest average rate of triple formation.  In fact, more tertiary systems than binary systems form in this case, in a ratio of $1.6  :  1$ as compared to $0.5  :  1$ for 100 M$_{\odot}$ simulations (Table~\ref{trpmass}).  Our artificial treatment of triples limits our ability to analyze reliably clusters with an IMBH $\ga$500 M$_{\odot}$.  In the 100-200 M$_{\odot}$ range, we do not find any conclusive pattern in the percent of triples with close orbits, or in the percent of mass transfer affected by triples.

\begin{deluxetable}{lrrrrr}
\tablecaption{Triple systems and IMBH companions \label{trpmass}}
\tablewidth{0pc}
\tablecolumns{6}
\tablehead{
  \colhead{IMBH Mass [M$_{\odot}$]}
& \colhead{50}
& \colhead{100}
& \colhead{150}
& \colhead{200}
& \colhead{500}
}

\startdata
MC Runs & 20 & 109 & 53 & 16 & 39 \\ & & & & & \\
$\langle {\rm N_{trip}} \rangle$ & 0.90 & 3.46 & 7.98 & 9.44 & 35.59 \\ & & & & & \\
$\langle {\rm N_{trip}/N_{bin}} \rangle$ & 0.38 & 0.50 & 0.75 & 0.81 & 1.58 \\ & & & & & \\
\% ${\rm trip_{close}}$ & 50.00 & 28.90 & 18.91 & 26.49 & 22.05 \\ & & & & & \\
\% MT$_{\rm aff}$ & 66.67 & 47.72 & 62.22 & 33.33 & 71.43 \\ & & & & & \\
\% MT$_{\rm close}$ & 0.00 & 10.24 & 13.33 & 6.67 & 20.41 \\ & & & & & \\
\% t$_{\rm MT}$ & 0.32 & 1.60 & 1.11 & 1.00 & 2.02 \\ & & & & & \\
\% t$^{\rm MT}_{\rm aff}$ & 3.13 & 21.07 & 41.82 & 25.16 & 67.08 \\ & & & & & \\
\% t$^{\rm MT}_{\rm close}$ & 0.00 & 4.96 & 7.57 & 0.00 & 27.14 \\
\enddata

\tablecomments{All simulations adopt cluster model A.  Abbreviations given in Table~\ref{trp}.}
\end{deluxetable}

\subsection{Cluster Models}

\begin{figure}
\epsscale{1.2}
\resizebox{\hsize}{!}{\includegraphics{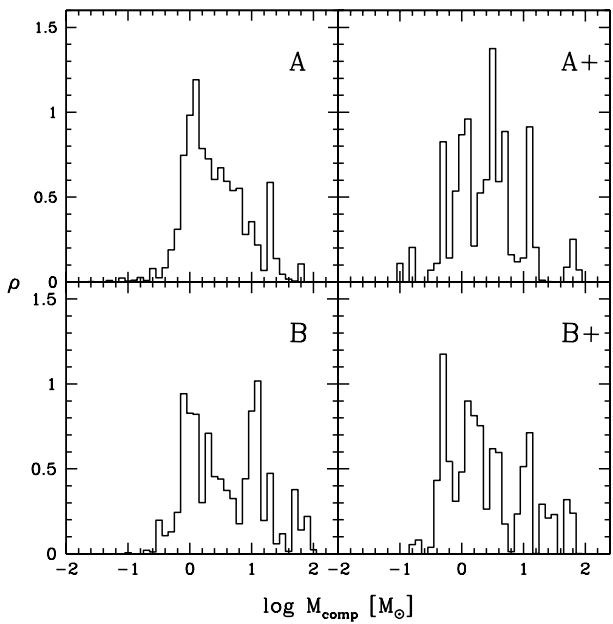}}
%\plotone{figure8.eps}
\caption{Probability density of companion mass at each timestep, shown for cluster models A (167 runs), B (102 runs), A+ (12 runs), and B+ (19 runs).\label{it_mdist}}
\end{figure}

\begin{deluxetable}{lccccc}
\tablecaption{ Binary companions of 100 M$_{\odot}$ IMBHs \label{lowic}}
\tablewidth{0pc}
\tablecolumns{6}
\tablehead{
  \colhead{Model}
& \colhead{A}
& \colhead{A+}   
& \colhead{B}
& \colhead{B+}
& \colhead{C}
}

\startdata

MC Runs                                  & 167 & 12 & 102 & 19 & 37 \\ & & & & & \\
$\langle {\rm N_{comp}} \rangle$         & 6.49 & 7.75  & 4.35 & 4.32 & 2.16 \\ & & & & & \\
$\langle {\rm N_{MS}} \rangle$           & 5.90 & 6.92  & 3.15 & 3.05 & 2.03 \\ & & & & & \\
$\langle {\rm N_{postMS}} \rangle$      & 0.47  & 0.50 & 0.59 & 0.58 & 0.13 \\ & & & & & \\
$\langle {\rm N^{MT}_{MS}} \rangle$      & 0.26  & 0.50 & 0.16 & 0.21 & 0.05 \\ & & & & & \\
$\langle {\rm N^{MT}_{postMS}} \rangle$ & 0.23  & 0.25 & 0.21 & 0.21 & 0.03 \\ & & & & & \\
\% MS$_{\rm MT}$                         & 4.36 & 7.23 & 4.97 & 6.90 & 2.67 \\ & & & & & \\
\% postMS$_{\rm MT}$                    & 50.00 & 50.00 & 35.00 & 36.36 & 20.00 \\ & & & & & \\
\% t$_{\rm comp}$                        & 57.63 & 61.43 & 40.35 & 39.82 & 34.71 \\ & & & & & \\
\% t$_{\rm MS}$                         & 40.87 & 48.90 & 18.88 & 23.19 & 32.35 \\ & & & & & \\
\% t$_{\rm postMS}$                    & 4.44 & 5.56 & 4.69 & 2.10 & 0.50 \\ & & & & & \\
\% t$^{\rm MT}_{\rm MS}$            & 2.81 & 1.43 & 0.54 & 0.17 & 0.52 \\ & & & & & \\
\% t$^{\rm MT}_{\rm postMS}$       & 0.11 & 0.05 & 0.04 & 0.07 & 0.01 \\

\enddata
\tablecomments{See Table~\ref{cm} for details of each model.  Models A+ and B+ are identical to A and B, but with 3BBF allowed.  Abbreviations given in Table~\ref{msrg}.}
\end{deluxetable}

To determine the dependence of our results on initial cluster conditions, we test two variations of our standard cluster model, as outlined in Table~\ref{cm}.  We use our standard IMBH mass of 100 M$_{\odot}$ for all these simulations. 

In model C, when a lower number density is used (n$_{\rm c} = 10^{4}$ pc$^{-3}$), the IMBH has a companion for only $\sim35\%$ of its evolution time, about half the companion time fraction calculated for the higher-density model (Table~\ref{lowic}).  The IMBH spends only a fraction of a percent of total simulation time with mass-transferring companions (0.53\%).  We conclude that clusters with our standard core number density, n$_{\rm c} = 10^{5}$ pc$^{-3}$, are much more likely to contain an accreting-IMBH ULX.

IMBH companions in clusters with a flattened IMF (model B) are slightly more massive, as expected, with most between $\sim1-10 {\rm M}_{\odot}$ (Fig.~\ref{it_mdist}).  The IMBH has a binary companion for a shorter time fraction (40\%) than in model A simulations (Table~\ref{lowic}).  A decrease is also seen in the mass transfer time fraction between the two models.  This could possibly result from having more massive stars in the core, due to the flatter IMF, that interact with and can disrupt IMBH binaries.  However, we caution again that the mass transfer time fraction for our standard model is dominated by a few outliers, so that the difference seen in model B may or may not be significant.

The time in which the model B clusters are observable as ULX sources is very short, as in model A clusters: a transient ULX forms for about 3.5\% of total mass transfer time, and persistent mass transfer at ULX luminosities occurs for about 7\% of mass transfer time. These imply a possible ULX time fraction of $< 0.05$\%. Furthermore, the behavior of triple systems in model B is qualitatively and quantitatively similar to that of triple systems in standard model clusters.

Small differences can be seen when 3BBF is allowed in the cluster.  3BBF enhances the rate at which massive objects can grow through collisions, since the hard formed binaries can collide with other formed binaries.  As a result, companions are a bit more massive in clusters with 3BBF (Fig.~\ref{it_mdist}).  The difference in the mass distributions is small, but a larger effect is seen in the mass transfer time fraction.  The time the IMBH spends with a mass-transferring companion decreases by more than half when 3BBF is included.  This phenomenon is probably similar to that which causes the disparity in mass transfer time between model A \& B clusters.  

Due to the 100\% initial binary fraction, we do not need 3BBF to form binaries.  Overall, we see that our cluster models are qualitatively similar with and without 3BBF, so we conclude that we can safely omit 3BBF from our simulations.

\section{Discussion}

We have undertaken the first detailed investigation of IMBH interactions in young ($\simeq 100$\,Myr), dense cluster cores, incorporating binary dynamics with full binary star evolution.  Specifically, we study the prevalence and characteristics of interacting binaries involving the central IMBH, and we examine and can better understand their role as potential ULX sources.  We use numerical cluster simulations that simplify some aspects of detailed dynamical evolution and allow us to focus on individual core stellar interactions while largely ignoring long-term cluster evolution. We analyze our results in terms of the frequency of IMBH binaries formed and their properties, the frequency of associated mass-transferring IMBH binaries, and the associated X-ray luminosities expected. With our cluster model A, we explore an IMBH mass range of 50-500 M$_{\odot}$, and we study two variations on this model as well (Table~\ref{cm}). Finally, we analyze the nature of triple systems formed as a possible influence on our results here and as a guide for designing future simulations.  

Our main results are summarized as follows:
\begin{itemize}
\item IMBHs more massive than $100$\,M$_{\odot}$ in model A clusters spend the majority ($>50$\%) of their evolution time with a binary companion. 
\item Mass-transferring IMBH companions, though, are relatively rare ($\simeq3\%$ of the IMBH evolution time) compared to the total time the IMBH spends with a companion. 
\item IMBHs spend more time in mass transfer phases with MS companions than with post-MS companions.
\item A much larger number fraction of post-MS companions than MS companions undergo mass transfer. 
\item In model A clusters, IMBH binaries are found to be persistent X-ray sources for about 90\% of the total mass transfer time, but the X-ray luminosity reaches ULX values (L$_{\rm X} > 10^{39}$\,erg\,s$^{-1}$) for only $\sim$2\% of the mass-transfer time.  For the rest of the mass-transfer time ($\simeq$10\%), IMBH binaries are found to be transients, which have short duty cycles. Given that the typical total mass-transfer time is $\simeq3\%$ of IMBH evolution time, these results amount to a vanishingly small observable-ULX time fraction ($<1\%$) over the lifetime of the young clusters we consider. 
\item Higher-mass IMBHs capture more companions throughout their evolution, making them more likely in a statistical sense to capture a companion that can overflow its Roche lobe.     
\item Model A clusters with a $50  {\rm M}_{\odot}$ IMBH and cluster model C have very few mass-transferring IMBH binaries.
\item In model B clusters, the IMBH spends less time with a companion (40\% of total time) and less time in mass transfer ($<1\%$ of total time) than in Model A clusters.
\item For all cluster models and IMBH masses explored here, the possible ULX time fraction of the cluster lifetime (100\,Myr) is found to be typically of $\sim 0.1$\% or even smaller. 
\end{itemize}

We note that results presented by \citet{hpza04} imply a higher incidence of possible ULX activity associated with MS companions to central IMBHs (typically with duration of $\gtrsim 10^7$\,yr). However, these calculations are based on analytical estimates of the possible tidal capture rates for an IMBH in a cluster of single MS stars and the subsequent {\em isolated} binary evolution once a MS companion is acquired. These binary evolution calculations are not integrated with any dynamical evolution (analytically or numerically) and, based on our results, we conclude that the assumption of isolated evolution is not realistic. Given the substantial differences in calculation methods, however, it is not possible to make direct and quantitative comparisons between the study by \citet{hpza04} and our study. We note that in our simulations we do allow for tidal captures (for details of implementation see Ivanova 2005), but they turn out to be a negligible contributor to the rate at which IMBHs acquire companions (exchanges into binaries is by far the dominant process).

Last, we recognize triple systems as one possibly limiting factor in our simulations.  Since triples are artificially disrupted as they form, we examine the influence of triples in terms of their initial orbital parameters and their coincidence with mass transfer phases. We find that while a large number of triple systems form per simulation, and while about $1/3\,-\,2/3$ of all 100-200 M$_{\odot}$ mass-transferring companions do coincide with triple system formation, only a small fraction of mass transfer phases are affected by triples with ``close'' orbits. Furthermore, two opposing phenomena related to triple breaking may cause some uncertainties.  Mass transfer may be prevented when potential Kozai evolution is artificially interrupted, or it may be artificially induced when release of the outer companion shrinks the inner binary.  We find that both effects are relevant for only small, roughly equivalent fractions of triples, causing a negligible net impact on our mass transfer results.  Therefore, we conclude that our treatment of triples in the regime of IMBH masses considered here does not influence our conclusions about the possibility of IMBH ULX formation. However, given the higher occurrence of triples in our $500$\,M$_\odot$ runs, the issues of multiples and cusp formation are most probably more important for more massive IMBHs ($\gtrsim 500$\,M$_\odot$) than those considered here. 

For these more massive IMBHs, a more physical treatement of multiple systems would in principle allow the remaining uncertainties regarding triples, multiples, and the ultimate formation of central cusps to be resolved. However, to examine ULX formation, treatment of the stellar evolution of multiple systems cannot be ignored (especially when it involves massive-star companions that lose mass in winds, experience core-collapse and can themselves disrupt orbits and alter dynamical evolution). Simulations that incorporate all of these effects are simply not possible at present. Instead, for our future explorations of the IMBH-ULX connection in dense clusters, we plan to implement simulations that would represent an improvement over the ones presented here. Such future calculations could differentiate between triple systems in which the inner binary may be disturbed and those in which the outer companion may be ignored.  In the latter case, the third companion would be kept in the system and its dynamical evolution alone could be simulated, while full stellar evolution and dynamics are calculated for the inner binary.

\acknowledgments
We thank Marc Freitag for valuable discussions on cluster dynamics and cluster properties before and after IMBH formation.  
This work is supported by a David and Lucile Packard Foundation
Fellowship in Science and Engineering grant, NASA ATP grant
NAG5-13236 and LTSA grant NAG5-13056 to V.\ Kalogera, NASA {\em Chandra} Theory Award NAS8-03060 to N.\ Ivanova, and NASA ATP grant NAG5-12044 to F.\ Rasio.

%\clearpage

\end{document}